\documentclass[12pt]{iopart}
\usepackage{graphicx}

\begin{document}

\title{High critical-current density and scaling of phase-slip processes in YBaCuO nanowires}

\author{G. Papari$^{1}$, F. Carillo$^{1}$, D. Stornaiuolo$^{2,3}$, L. Longobardi$^{3}$, F. Beltram$^{1}$ and F. Tafuri$^{3,2}$}

 \address{$^{1}$  NEST Scuola Normale Superiore and
CNR-Istituto Nanoscienze, Piazza San Silvestro 12, I-56127
Pisa,Italy} \address{$^{2}$ CNR-SPIN UOS Napoli, Complesso
Universitario di Monte Sant'Angelo via Cinthia, 80126,
Napoli,Italy} \address{$^{3}$ Dip. Ingegneria dell'Informazione
Seconda Universit\`{a} di Napoli, Aversa (CE) and CNR-SPIN,
Napoli, Italy}\ead{paolo.papari@fys.kuleuven.be}

\date{\today}

\begin{abstract}

YBaCuO nanowires were reproducibly fabricated down to widths of 50
nm. A Au/Ti cap layer on YBCO yielded high electrical performance
up to temperatures above 80 K in single nanowires. Critical
current density of tens of MA/cm$^2$ at T = 4.2 K and of 10
MA/cm$^2$ at 77 K were achieved that survive in high magnetic
fields. Phase-slip processes were tuned by choosing the size of
the nanochannels and the intensity of the applied external
magnetic field. Data indicate that YBCO nanowires are rather
attractive system for the fabrication of efficient sensors,
supporting the notion of futuristic THz devices.
\end{abstract}

\maketitle

\section{Introduction}

Superconducting nanostructures operating at the liquid nitrogen
temperature are very attractive for several applications  in
superconducting electronics. In particular high
critical-temperature superconductor (HTS) nanowires not only have
the advantage of high temperature operation, but in principle can
be functionally scaled to smaller sizes \cite{wei,caltech} thanks
to their extremely short coherence lengths $\xi_0$. Other
intrinsic properties of these materials can play a relevant role
too. For instance, their characteristic fast relaxation times
\cite{sobol} offer higher counting rates in photo-detection
experiments when compared to traditional superconductors
\cite{goltsman}.

In this context reliable top-down fabrication techniques allowing
a more direct integration of the HTS nanostructures into hybrid
systems are of much interest. Here we analyze the electrical
properties of a number of nanofabricated wires down to 50 nm in
width. Our measurements show very good performance for single YBCO
nanowires both in terms of critical temperature ($T_C$) and
critical-current density ($J_C$) with excellent potential for sensor
applications. Critical-current densities up to 70 MA/cm$^2$ at T =
4.2 K and 10 MA/cm$^2$ at 77 K were achieved on YBCO capped by a
thin protective layer of Au/Ti. These are among the best results
reported for single nanowires
\cite{bonetti,storici,herbstritt,curtz} (see below). Contrary to
other works \cite{caltech}, where even smaller sizes were achieved
for $arrays$ of nanowires, in this case we isolated single
nanowires, thus responding to a much wider range of requirements
for circuit design.

\section{Experimental details}

We have employed YBCO thin films produced on various substrates
through different deposition techniques (sputtering and reactive
thermal co-evaporation \cite{kinder}). For the aims of this work
of demonstrating a reliable top-down technology to obtain single
YBCO nanowires down to a width of 50 nm and the utility of using a
Au cap layer, the properties of the thin films to privilege are
rms roughness as low as possible, and absence of macroscopic
impurities in the film. The initial values of $T_C$ and $J_C$ of
unpatterned films are relevant but not crucial. In order to
perform systematic experiments and reliably compare nanostructures
realized on different samples, we have employed thin films
produced on large areas by co-evaporation technique by Theva
\cite{theva}. From the same wafer, it is possible to cut 20
samples of size of 5 mm x 5 mm with the same properties. In this
work we focus on c-axis YBCO thin films (50 nm thick), grown on a
$CeO_2$ seed layer (40 nm thick), and protected by a 20 nm thick
Au layer \cite{theva}. The substrate is a large wafer of
Yttria-stabilized zirconia (YSZ). The values of the original
critical temperature and of the critical current density are high
($T_C$ $\simeq$86.5 K and $J_C$= 2.2 $MA/cm^2$ at T= 77 K) but not
optimized to the highest possible values. The key-steps of the
nanotechnology approach are based on the use of a 30 nm thick Ti
mask patterned through e-beam lithography (ZEISS MERLIN), and on a
low energy milling procedure keeping the sample at a temperature
of about -150$^oC$. Most of the procedure is described in detail
in \cite{IEEE}. All nanobridges described in this work are 1
$\mu$m long. A comparative study of bare and capped YBCO nanowires
on samples produced in the same deposition run and only differing
by the last fabrication step (bare nanowires undergo an additional
ion milling step in order to remove the Au layer) sheds light on
the role of Au layer protecting the nanowires. The Au/Ti layer
makes easier the integration of the YBCO component into hybrid
structures because of its better compatibility with other
materials as opposed to HTS. If only genuine superconducting
properties are required, the YBCO capped nanowires can perfectly
replace YBCO nanowires for most functions. Nano-fabrication
techniques are not restricted, for the resolutions specified
above, to any specific choice of the substrates usually employed
for HTS thin films, or thin film deposition conditions, as long as
surface roughness is below some thresholds, for instance lower
than 5nm on a total thickness of 50 nm.


Measurements were performed in liquid helium using
a four probe configuration. The set up employed is equipped with
two stages of cold filters (copper powder and RC Pi filters with a
cut-off frequency of the order of a few GHz and 160 MHz
respectively).

\begin{figure}[t]
    \begin{center}
        \includegraphics[width=0.6\textwidth]{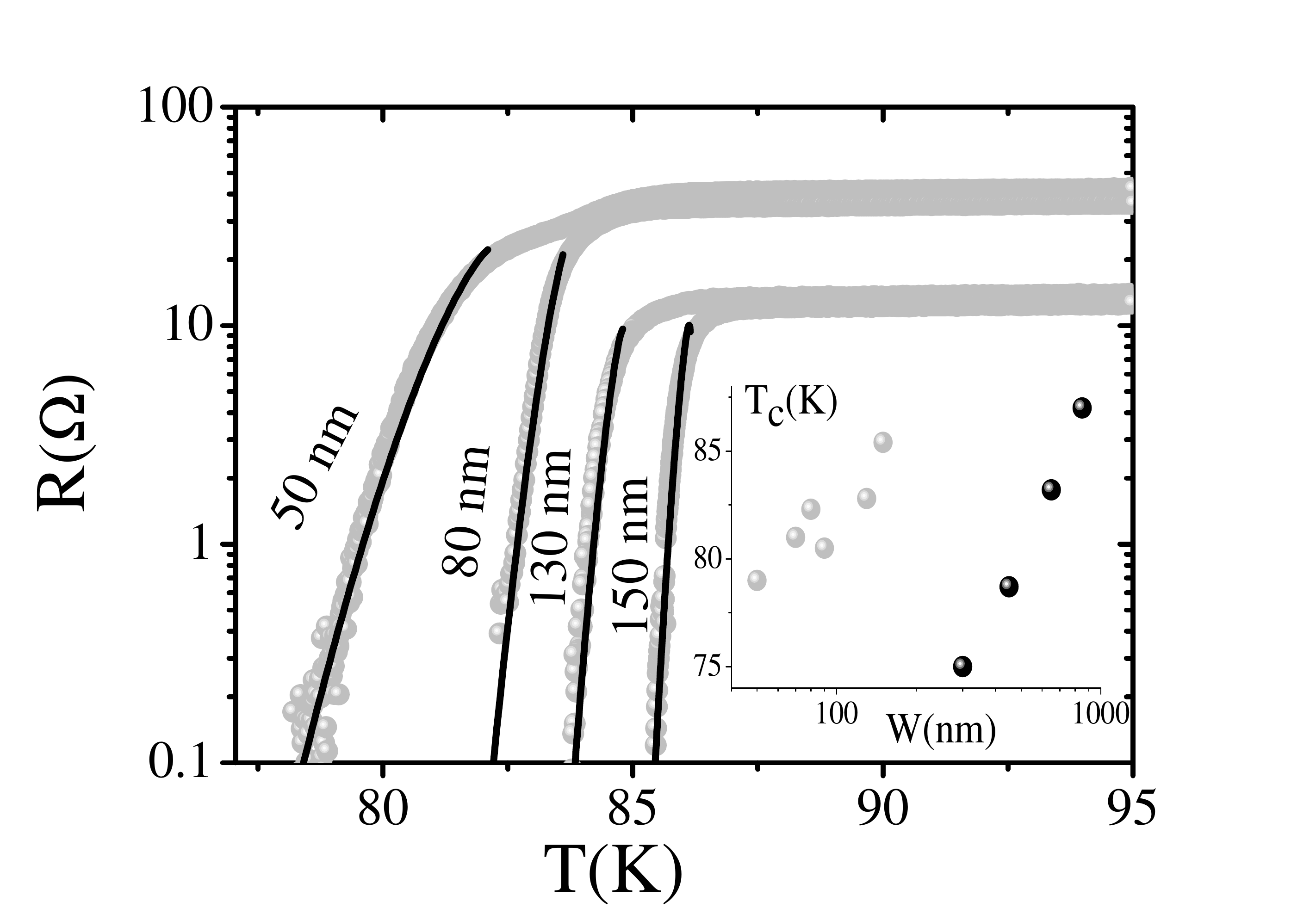}
    \end{center}
    \caption{ Resistive transitions of YBCO/Au/Ti
nanowires of different widths indicated in the figure. The dashed
lines are the fits achieved using the LAMH theory (see text).
Inset: Dependence of the critical temperatures on the nanowire
width. The grey dots refer to the T$_c$'s of YBCO/Au/Ti samples
while the black points to  bare YBCO nanowires.}
\end{figure}

\begin{figure}[t]
    \begin{center}
        \includegraphics[ width=0.7\textwidth]{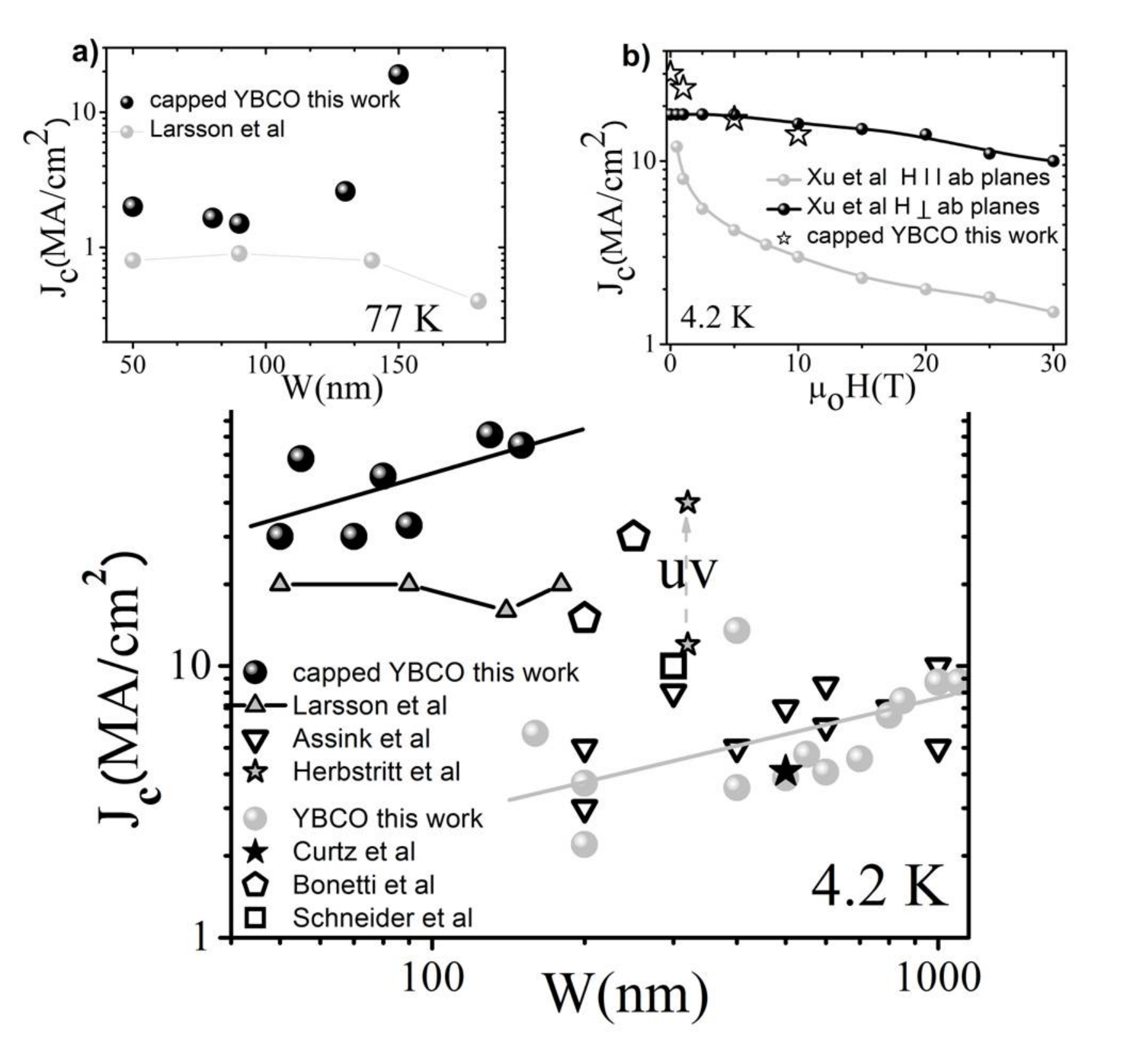}
    \end{center}
    \caption{Values of critical current density $J_C$ measured
 at 4.2K versus channel width (W). Data from this work (grey circles for YBCO nanowires,
 black circles for YBCO/Au/Ti nanowires) are compared with results available in literature
 \cite{bonetti,storici,curtz,herbstritt}. Uv stands for the increase of J$_C$ obtained by
ozone treatment \cite{herbstritt}.   Inset a): $J_C$ vs W of the
YBCO/Au/Ti nanowires measured at 77K are compared with
\cite{storici}. Inset b):  J$_c$ of the 50 nm wide nanochannel
(measured  at T= 4.2 K) at different values of  the magnetic field
H are compared with typical data from literature \cite{jcvsB}.}
\end{figure}

\section{Results and Discussions}

The critical temperature of various representative devices is
reported as a function of wire width in the inset of Fig. 1. $T_C$
scales linearly for both types of bridges. For the capped YBCO,
$T_C$ decreases from about 86 K for unpatterned films to about 85
(78) K for nanowires of width w= 140 (50) nm respectively. For the
bare YBCO samples, $T_C$ of 1$\mu$m wide bridges is basically the
same of the unpatterned films and linearly decreases down to 75 K
for nanowires of width w= 300 nm. Therefore the 'critical' region
is for widths below about 1 $\mu$m. Superconductivity is quite
robust also for widths less than 200 nm, but the properties of
devices are less reproducible.


Not surprisingly, the bare YBCO thin films are more sensitive
to ion milling and processing details. Figure 1 shows resistance
vs temperature curves for the capped YBCO nanowires. Both bare
and capped YBCO data can be well fitted in terms of the
thermally-activated phase-slip model, i.e. the Langer- Ambegaokar-
McCumber - Halperin (LAMH) model \cite{lahm} commonly used to
describe the broadening of R(T) in superconducting nanostructures
\cite{bezryadin}. The total resistance of the wire is commonly
expressed as the parallel \cite{bezryadin} between the resistance
activated by phase-slip processes $R_{LAMH}$
\begin{equation}
R_{LAMH}=R_{Q}\frac{\hbar\Omega}{k_{B} T}\exp\left(-\frac{\Delta
F}{k_{B}T}\right)\label{RLAHM}
\end{equation}
and the normal resistance of quasiparticles
$R_N(T)=R_N\exp\left(-\Delta F(T)/k_B T\right)$ \cite{little}:
$R(T)^{-1}=R_{LAMH}(T)^{-1}+R_N^{-1}(T)$. Here $
\Omega=\left(L/\xi\right)\sqrt{\left(\Delta F/k_B
T\right)}\left(1/\tau_{GL}\right)$ is the phase slipping rate,
$\tau_{GL}=\left(\pi\hbar/k_B\,\left(T_C-T\right)\right) $,
$R_Q=h/4e^2$ is the quantum resistance constant, L is the sample
length, $k_B$ is
the Boltzmann constant, 
and $\Delta F$ is the free energy where the superconducting phase
"falls off". $\Delta F$ at T=0 K can be expressed as $8\sqrt{2} /
(48 \pi^2) A(\phi_0^2/(\mu_0 \lambda^2 \xi_0 ))$ (where A and
$\lambda$ are the cross section and the London penetration depth
respectively) in analogy with other experiments on HTS nanowires
\cite{caltech}. The extracted values of the fitting parameters are
$\xi_0 \sim 2 nm$ and of  $\lambda \sim 300 nm$ which, are quite
consistent with typical values of the cuprates \cite{sudbo}.
$\Delta$F is for all curves about $10^5$ K.

Measured critical-current density values are reported as a
function of the width (w) for T = 4.2 K in Fig. 2 and compared
with results available in literature
\cite{bonetti,storici,herbstritt,curtz}. $J_C$ in capped YBCO
nanowires is higher than 30 MA/cm$^2$ (for w = 50 nm) and reaches
a maximum of 70 MA/cm$^2$ for W = 140 nm. These are among the
highest values reported \cite{bonetti,storici,curtz,herbstritt}
and only a few times lower than the theoretical depairing limit
(300 MA/cm$^2$) \cite{lang}. $J_C$ values at T = 77 K are reported
in Fig. 2 inset a) for YBCO/Au/Ti nanowires and range from 2
MA/cm$^2$ for w = 50 nm to 20 MA/cm$^2$ for W = 140 nm. The better
performances offered by capped nanowires when compared with those
of bare YBCO nanowires of this and previous works \cite{storici},
reveal that the protecting Au layer is key for enhancing the
quality of HTS nanowires. We believe that there are margins to
further improve $J_C$ performances especially at 77 K by employing
films with higher values of the initial $T_C$.

Nanowires keep good transport properties also when an external
magnetic field is applied. For instance, in the extreme limit of
50 nm, a magnetic field of 10 T reduces $T_C$ by less than 10 \%,
while $J_C$ at 4.2 K is only halved. Superconductivity in 50 nm
wide strips survives up to H = 1 T at 77 K. Figure 2 inset b)
allows a comparison of the dependence of $J_C$ on magnetic field
at T = 4.2 K \cite{jcvsB} in the present samples with other
results and confirms the robustness against an externally applied
magnetic field. The use of HTS for power applications is a field
of research of great relevance and a 'smart' control of pinning
centers is the key to increase $J_C$. In macroscopic samples this
role is usually played by the introduction of nanodefects
\cite{correnticritiche,correnticritiche2}, while in nanowires by
the detailed structure of the edge barrier \cite{tahara}.


\begin{figure}[t]
    \begin{center}
        \includegraphics[width=0.5\textwidth]{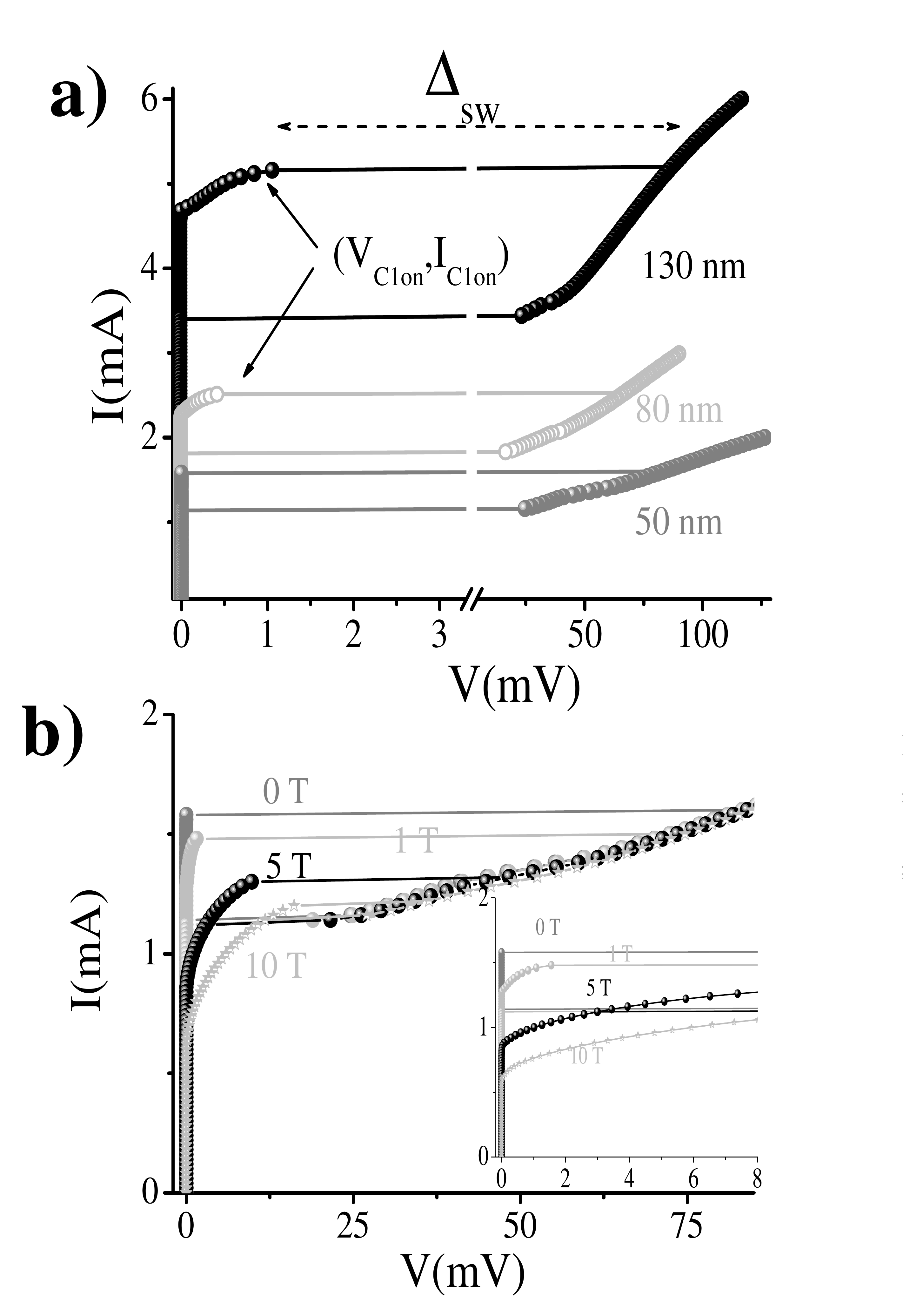}
    \end{center}
    \caption{  a) I-V curves at T = 4.2 K for YBCO/Au/Ti
nanowires of widths 130 nm, 80 nm and 50 nm, respectively. b) For
the 50 nm wide bridge, I-V curves are reported for different
values of the magnetic field (H=0,1 T, 5 T and 10 T). }
\end{figure}

Figure 3a shows the experimental current-voltage (I-V)
characteristics measured at T = 4.2 K for YBCO/Au/Ti nanowires
130, 80 and 50 nm wide. The distinctive feature is the presence of
steps accompanied by a very large hysteresis that exceeds 30\% of
the total current. Similar features were reported by several
author in experiments involving nanostructures  \cite{bezryadin}
and HTS nanowires  \cite{caltech,mikheenko,antognazza}. These
observations were interpreted in terms of phenomena driving the
superconducting channel to the normal state, i.e. the development
of phase-slip centers (PSC) or normal hot spots (HP)
\cite{skocpol}. Because of their low carrier density and thus low
superfluid rigidity, HTS systems are inherently prone to phase
fluctuations and therefore more susceptible to PSC in reduced
dimensions \cite{blatter,bezryadin}. I-V curves reported in Fig. 3
show that the amplitude and the position of the steps depends on
the size of the nanochannels and on the magnetic field. By
increasing the width of the channel, the switch occurs at higher
voltage. This is consistent with the idea that in wider samples
the transition to the fast vortex motion that leads to the step
structure occurs at a larger current \cite{vodolazov}. As shown in
Fig. 3b, an increase in H moves V$_{1on}$ to higher voltages to a
maximum of about 15 mV for H = 10 T. The voltage jump
$\Delta_{SW}$ is also reduced from $\Delta_{SW}$= 80 mV at H = 0 T
to about 10 mV at H = 10 T.

From the value of the critical voltage V$_{1on}$=$v_C$HL \cite{larkin,vodolazov}
it is possible to estimate the vortex critical velocity $v_C$=$ 10^3$ m/s (at H= 1 T),
and hence the inelastic electron-electron scattering time $\tau_S\approx0.6ns$
on the basis of the expression $\tau_S=D[14\zeta(3)]^{0.5}\left(1-T/T_C\right)^{0.5}/\pi v_C^2$ \cite{huebener}, where D is the quasiparticle diffusion
constant and $\zeta{(3)}\sim1.2$ . This value is in agreement with other experiments \cite{huebener} and estimates \cite{vodolazov}.


The scaling of the I-V curves and of the steps with the width of
the nano-channels and the influence of H on the I-Vs reflects the
presence of phase slip lines (PSL) \cite{sivakov,vodolazov}. PSL
are a 2-dimensional analogue of PSC and reveal some kind of phase
transition in the vortex lattice at the instability point with
regions of fast and slow vortex motion \cite{sivakov,vodolazov}.
When the width of the bridges is reduced to w = 50 nm, the step
position (V$_{1on}$ = 0) and the shape of the I-V indicate a
transition from PSL to the more classical 1-dim PSC (at H = 0 T).
In the narrowest nanowire, not more than one line of vortices at
low fields can be hosted. When the magnetic field is increased,
the size of inter-vortex distance $a$ ($a = \sqrt{\Phi_o/H}$) is
decreased (for instance a = 20 nm at H = 5 T), and the entrance of
additional rows of vortices is favored, with a consequent
transition to PSL. This is documented by the steps at finite
voltage in the presence of magnetic field (V$_{1on} >$ 0). Vortex
motion is characterized by oscillatory behavior of flux lines
which cross the sample resulting in an emission of electromagnetic
radiation $\nu=v_c/a$ that is on the order of $0.2\times10^{11}$Hz
at H=1T.

Studies on the dynamics of kinematic vortex-antivortex lines
confirm that their individual velocity can be manipulated by
applying magnetic field and current \cite{Berdiyorov}, supporting
the notion of futuristic THz devices.



\section{Conclusions}

In conclusion single YBCO nanowires were fabricated through a
Au/Ti cap layer to reach very high values of the critical current
density also at 77 K that is very robust against an externally
applied magnetic field H. The detailed structures of the I-V
curves of these single nanowires can be tuned by choosing the size
of the bridge and by setting the magnetic field intensity.
Threshold mechanisms associated to phase slips processes and
visible as steps in I-Vs pave the way to HTS nanowires as sensors
for photodetection experiments.

\ack{ The authors would like to thank A. Barone, F. Lombardi, D.
Montemurro, V. Moshchalkov, V. Piazza, S. Roddaro and R.
Wordenweber for fruitful discussions. This work has been partially
supported by ESF "MIDAS" project, by a Marie Curie International
Reintegration Grant No. 248933 "hybMQC" and by MIUR PRIN 2009
under the project "SuFET based on nanowires and HTS"}

\end{document}